# Note on Breakup Densities in Fragmentation

V. E. Viola
IUCF and Department of Chemistry
Indiana University
Bloomington, IN 47408

In [1] the breakup density of hot nuclei as a function of excitation energy $E^*/A$ was derived from analysis of kinetic-energy spectra for intermediate-mass fragments ( IMF: $2<Z\sim20$ ) emitted in light-ion induced reactions on $^{197}$Au nuclei. The results indicated a breakup density that decreases from normal density $\rho_o$ at excitation energies below $E^*/A \sim 2$ MeV to a value of $\rho/\rho_o \sim 0.3$ for $E^*/A \sim 5$ MeV and above. This analysis has been questioned [2,3] on the grounds that a parameterization based on a sequential emission model [4] was used to fit the spectra, seemingly in conflict with the observation that the time scale is nearly simultaneous above $E^*/A \sim 3$ MeV. The purpose of this note is to elaborate on the assumptions that were employed in [1] and to stress that the sequential emission formalism of [4] served only as a spectral fitting function, independent of time scale.

The analysis in [1] was prompted by the observed systematic decrease in the IMF spectral "Coulomb" peaks with increasing excitation energy [6,7], contrary to the expected increase with the higher temperature of the emitting system. Two plausible explanations for the observed behavior are (1) a decrease in source charge due to fast cascade/preequilibrium charge loss prior to thermalization, and (2) a decrease in density due to expansion/dilution of the emitting source. The source charge can be corrected from the data, since the nonequilibrium particles were measured experimentally [8]. The extent of expansion/dilution is reflected by the separation distance between a fragment and the Coulomb field generated by the ensemble of remaining particles. Using Coulomb's law, this distance can be derived from the fragment kinetic energy spectra, as described below.

Our analysis is first-order and descriptive of an average process. For high multiplicity events the measured spatial patterns of the fragments involve a broad distribution of configurations that can only be addressed collectively. In addition, an effort was made to favor maximum breakup densities whenever ambiguities in the procedure occurred. The principal assumptions for the disintegrating system are as follows:

- The average breakup configuration is isotropic, as has been indicated by a sphericity/coplanarity analysis [9]. Fluctuations in the event topology are assumed to be reflected in the widths of the spectra.
- Only equilibrium-like particles are included in the fitting procedure and nonequilibrium particles are subtracted from the target charge to determine the source charge. (This procedure agrees well with the results of the EOS collaboration [10]).
- Fragment acceleration is governed only by the average Coulomb field between the an IMF and the residual system; i.e. there is no significant radial flow [11], although this possibility is also addressed in [1].
- From a knowledge of the IMF kinetic energy distributions and the charges of the IMF and the residue, the average separation distance can be extracted, defining the breakup volume for a spherical system.

Thus, the problem is reduced to one of relating the IMF kinetic energy spectra to some average separation distance and finding an appropriate normalization to $\rho_o$.

The spectral analysis in [1] is analogous to that employed in the derivation of fission-fragment kinetic-energy systematics, from which fragment-fragment charge separation distances can be derived [12]. The principal difference is that the functional dependence in [1] is on $E^*/A$ instead of $Z^2/A^{1/3}$ as in [12]. At moderate excitation energies the mass division in fission is symmetric, leading to fragment kinetic energy distributions that are Gaussian-like and thus uniquely characterized. The effect of excitation energy is reflected only in the widths of the distributions [13]. In contrast, IMF emission is asymmetric and the kinetic-energy distributions are Maxwellian in shape – whether in binary emission or multifragmentation. This leads to several possible ways to describe the shifts in the spectral peaks.

In order to describe the spectral shifts, the choice of the Coulomb parameter from the Moretto model [4], as modified by Kwiatkowski [14], was dictated by practical considerations. Most important was that the Coulomb parameters had been previously obtained in moving-source fits that were performed in order to separate equilibrium and nonequilibrium emissions, determine source velocities, and compare slope temperatures with Fermi-gas expectations. Specific features included the following:

- When the separation distances based on fission fragment kinetic-energy systematics are used for the radius parameter in the model, excellent fits to the spectra for binary emission at normal density are obtained [7]. This assumption provided our normalization to normal density.
- The program contained the kinematic factors required to assure consistency in the source frame transformations.
- The model allowed for the observed broadening of the spectral widths with increasing $E^*/A$ due to thermal and fragment deformation effects.
- Corrections to the source charge were applied to account for mass loss due to nonequilibrium emission.

This approach allowed us to fit hundreds of spectra in order to obtain systematic Coulomb parameters, source velocities and slope temperatures as a function of both IMF charge and $E^*/A$. Nowhere in the analysis is the assumption of sequential emission used. Although much earlier we pointed out the qualitative link between the decreasing Coulomb parameters and the breakup density, the quantitative analysis in [1] was a light that snapped on much later.

It should be stressed that we could have chosen any number of metrics for fitting the IMF kinetic-energy spectra -- e.g. the most probable, average, or centroid at FWHM of the kinetic energy distributions. The results would not be significantly different than in [1], although more poorly defined. Some concern exists that the breakup densities in [1] are somewhat lower than those derived from caloric curves by Natowitz [15], derived largely from heavy-ion data. Possible sources for the difference may reside in the compression/expansion energy and/or rotational effects that may enhance breakup at a somewhat higher density than in light-ion-induced reactions. In addition, the reaction dynamics of GeV light-ion reactions indicate that the nucleus is formed in a relatively dilute state due to multiple knockout reactions during the fast cascade that initiates the energy deposition process [16].

One final note on this analysis. The densities quoted in [1] are based on an average of values for individual IMFs from $Z = 4-10$. Closer examination of results for individual IMFs reveals a trend in which the breakup density increases as the IMF charge decreases, such that for He ions the breakup density is close to normal density. However, within experimental errors, this effect is only suggestive.